\newcommand{\op}[1]{\hat {#1}}
\newcommand{\Tr}{\mathop{\mathrm{Tr}}\nolimits}

\documentclass[pra,twocolumn,showpacs,amsmath,amssymb,eqsecnum,superscriptaddress]{revtex4-1}
\usepackage{graphicx,bm,color,mathptmx,hyperref}

\begin{document}

\title{Orbital angular momentum from marginals of quadrature distributions}

\author{L.~L.~S\'anchez-Soto} 
\affiliation{Departamento de \'Optica,  Facultad de F\'{\i}sica, 
Universidad Complutense, 28040~Madrid,  Spain}
\affiliation{Max-Planck-Institut f\"ur die Physik des Lichts, 
G\"{u}nther-Scharowsky-Stra{\ss}e 1, Bau 24, 
91058 Erlangen,  Germany}
\affiliation{Department f\"{u}r Physik, Universit\"{a}t Erlangen-N\"{u}rnberg,
Staudtstra{\ss}e 7, Bau 2, 91058 Erlangen, Germany}

\author{A. B. Klimov} 
\affiliation{ Departamento de F\'{\i}sica,  Universidad de  Guadalajara, 
44420~Guadalajara, Jalisco, Mexico}

\author{P. de la Hoz}
\affiliation{Departamento de \'Optica,  Facultad de F\'{\i}sica, 
Universidad Complutense, 28040~Madrid,  Spain}

\author{I. Rigas} 
\affiliation{Departamento de \'Optica, Facultad de F\'{\i}sica, 
Universidad Complutense, 28040~Madrid, Spain}
\affiliation{Max-Planck-Institut f\"ur die Physik des Lichts, 
G\"{u}nther-Scharowsky-Stra{\ss}e 1, Bau 24, 
91058 Erlangen,  Germany}

\author{J. \v{R}eh\'{a}\v{c}ek}
\affiliation{Department of Optics,
Palack\'{y} University, 17. listopadu 12, 
746 01 Olomouc, Czech Republic}

\author{Z. Hradil}
\affiliation{Department of Optics,
Palack\'{y} University, 17. listopadu 12,
746 01 Olomouc, Czech Republic}

\author{G.~Leuchs}
\affiliation{Max-Planck-Institut f\"ur die Physik des Lichts, 
G\"{u}nther-Scharowsky-Stra{\ss}e 1, Bau 24, 
91058 Erlangen,  Germany}
\affiliation{Department f\"{u}r Physik, Universit\"{a}t Erlangen-N\"{u}rnberg,
Staudtstra{\ss}e 7, Bau 2, 91058 Erlangen, Germany}

\begin{abstract}
  We set forth a method to analyze the orbital angular momentum of a
  light field. Instead of using the canonical formalism for the
  conjugate pair angle-angular momentum, we model this latter variable
  by the superposition of two independent harmonic oscillators along
  two orthogonal axes. By describing each oscillator by a standard
  Wigner function, we derive, via a consistent change of variables, a
  comprehensive picture of the orbital angular momentum. We compare
  with previous approaches and show how this method works in some
  relevant examples.
\end{abstract}


\date{\today}

\maketitle

\section{Introduction}

The term vortex is commonly used to designate a region of concentrated
rotation in a flow, such as an eddy, a whirlpool, or the depression at
the center of a whirling body of air or water. Naturally occurring
vortices include hurricanes, tornadoes, waterspouts, and dust
devils~\cite{Childs:2011fk}. Yet vortices can also be created in many
different media: they manifest in plasmas~\cite{Mikhailovskii:1987ys},
superfluids~\cite{Salomaa:1987ly}, ferromagnets~\cite{Hubert:1998ve},
acoustical waves~\cite{Hefner:1999qf}, quantum Hall
fluids~\cite{Ezawa:2000dq}, Bose-Einstein
condensates~\cite{Pitaevskii:2003kx}, and electron wave
packets~\cite{Bliokh:2007cr}, to cite only a few relevant
examples. This points to the ubiquity of this phenomenon and reveals
a growing interest in these singularities.

The case of optical vortices deserves a special
mention~\cite{Torres:2011vn}. An optical vortex is a beam of light
exhibiting a pure screw phase dislocation along the propagation axis;
i.e., an azimuthal phase dependence $\exp ( i \ell \phi)$. The number
$\ell$ plays the role of a topological charge: the phase changes its
value in $\ell$ cycles of $2 \pi$ in any closed circuit about the
axis, while the amplitude is zero there.

One of the most interesting properties of vortices is that they carry
orbital angular momentum (OAM): the integer $\ell$ can be seen as the
eigenvalue of the OAM operator and its sign defines the helicity or
direction of rotation. Indeed, the OAM of such a field can be
easily manipulated and transferred, which opens many experimental
perspectives, such as optical tweezers and
spanners~\cite{Padgett:2010tg}, as well as potential
astronomical~\cite{Elias:2008hc} and communication
applications~\cite{Wang:2012bs}.

The fact that individual photons also carry OAM presents the most
exciting possibilities for using this variable in the quantum domain,
and a number of uses has already been demonstrated~\cite{Mair:2001fv,
  Molina:2004dz,Oemrawsingh:2004fu,Marrucci:2006kl,Molina:2007qa}.

In quantum theory, the operator representing the OAM has an unbounded
spectrum that includes positive and negative integers. Accordingly,
its conjugate variable, the azimuthal angle, might be expected to
be represented by a \textit{bona fide} operator. Periodicity, however,
brings out subtleties that have triggered long and heated
discussions~\cite{Lynch:1995cq,Perinova:1998xq,Luis:2000fk}.

Here, we look at this issue from a phase-space perspective.
Such an approach was introduced in the very early days of quantum
theory to avoid some of the troubles arising in the abstract
Hilbert-space formulation. The pioneering works of
Weyl~\cite{Weyl:1928}, Wigner~\cite{Wigner:1932uq}, and
Moyal~\cite{Moyal:1949fk} paved thus the way to formally picturing the
quantum world as a statistical theory on phase
space~\cite{Schroek:1996fv,Schleich:2001hc,QMPS:2005mi}.

In few words, the key idea is to look for a mapping relating operators
(in Hilbert space) to functions (in phase space).  For the conjugate
pair angle-OAM, the phase space is the discrete cylinder $\mathcal{S}
\times \mathbb{Z}$ ($\mathcal{S}$ denotes the unit circle associated
with the angle, while $\mathbb{Z}$ are the integers labeling OAM). It
seems natural to work out a Wigner function (or any other
quasiprobability) therein.  A pioneer attempt in that direction was
made by Mukunda~\cite{Mukunda:1979uq,*Mukunda:2005kx}; his work was
subsequently reelaborated and developed in a variety of directions by
other authors~\cite{Bizarro:1994vn,Vourdas:1996ys,Nieto:1998cr,
  Ruzzi:2002rt,Zhang:2003zr,Kakazu:2006fr,Kowalski:1996mz,Gonzalez:1998gf,
  Ohnuki:1993ul,Hall:2002pd,Ruzzi:2006dq,Rigas:2010kx,Rigas:2011by}.

However, one might properly argue that in such a --correct-- way of
proceeding one is overlooking significant information about the
transverse distribution.  This means, for example, that, using
cylindrical coordinates, all the states $\Psi_{\ell} (r, \varphi) =
A_{\ell} (r) \exp(i \ell \varphi)$ represent eigenstates of the
angular momentum, irrespective of the form of the amplitudes $
A_{\ell} (r)$.  A similar problem arises in the description of
spinlike systems over the Bloch sphere: one disregards in this way
fluctuations in the number of particles, because a sphere of fixed
radius cannot accommodate those fluctuations. To bypass this drawback
one needs to include the whole Bloch space that can be envisioned as
foliated in a set of nested spheres with radii proportional to the
different number of particules that contribute to the state.

Below, we propose an alternative road and derive phase-space
distributions via suitable marginals of distributions for field
quadratures, once we remove the degrees of freedom irrelevant for the
specification of the problem. The same ideas have been used also to
study quantum polarization properties~\cite{Luis:2005lh,
  Klimov:2006ff}.  Perhaps, this provides the most down-to-earth
approach to the problem at hand, since the quadrature distributions
can be determined by very simple experimental
procedures~\cite{Lvovsky:2009dp}.  This widespread measurability does
not hold for the Wigner functions on the cylinder: the proposals for
their practical reconstruction are rather 
cumbersome~\cite{Rigas:2008nx,*Rehacek:2010fk} and lack the simple and
intuitive picture provided by schemes measuring quadrature
distributions.

The plan of this paper is as follows. In Sec.~\ref{Sec:1DHO} we
concisely sketch the phase-space fundamentals for a single harmonic
oscillator. In Sec.~\ref{Sec:2DHO} we start from two kinematical
independent orthogonal oscillators and express the resulting Wigner
function in cylindrical coordinates. By eliminating an unessential
variable (the radial momentum), we get a well-behaved distribution
that gives complete information, not only on the pair angle-OAM, but
also on the radial distribution. We apply the resulting Wigner
function to some relevant states in Sec.~\ref{Sec:ex}, and conclude
that it constitues a most suitable tool to deal with this problem.

\section{Phase-space picture of a one-dimensional harmonic oscillator}
\label{Sec:1DHO}

To keep the discussion as self-contained as possible, we first boil
down the rudiments of the phase-space formalism for a harmonic oscillator
that we shall need later on.

The relevant dynamical observables are the conjugate coordinate and
momentum operators $\op{x}$ and $\op{p}$, with canonical commutation
relation (with $\hbar = 1$ throughout)
\begin{equation}
  \label{eq:HWcom}
  [ \op{x}, \op{p} ] = i \, \op{\openone} \, ,
\end{equation}
so that they are the generators of the Heisenberg-Weyl
algebra~\cite{Binz:2008oq}.  Ubiquitous and profound, this algebra has
become the hallmark of noncommutativity in quantum theory. The
classical phase space is here the plane $\mathbb{R}^{2}$.

Sometimes, it is advantageous to use instead complex amplitudes
represented  by the annihilation and creation operators
\begin{equation}
  \label{eq:aadag}
  \op{a} = \frac{1}{\sqrt{2}} (\op{x} + i \op{p} )  \, ,
  \qquad \qquad
  \op{a}^{\dagger} = \frac{1}{\sqrt{2}} (\op{x} - i \op{p} ) \, ,
\end{equation}
in terms of which the commutation relation (\ref{eq:HWcom})
turns out to be $ [\op{a}, \op{a}^\dagger ] = \op{\openone}$.

A pivotal role will be played in what follows by the unitary 
\begin{equation}
  \label{eq:HWDisp1}
  \op{D} (x, p) =   \exp[i ( p  \op{x} -  x \op{p})] \, ,
\end{equation}
which is called the displacement operator for it displaces a localized
state by $(x, p) \in \mathbb{R}^{2}$.  The Fourier transform of $\op{D} (x, p)$
\begin{equation}
  \label{eq:HWkernelDef}
  \op{w} (x, p) = \frac{1}{(2\pi)^2} \int_{\mathbb{R}^2} 
  \exp[-i(p x^\prime - x p^\prime)] \, \op{D} (x^\prime, p^\prime) \, 
  dx^\prime dp^\prime \, ,
\end{equation}
is an instance of a Stratonovich-Weyl quantizer~\cite{Stratonovich:1956kx}.
One can check that the operators $\op{w} (x, p)$ are a complete
trace-orthonormal set that transforms properly under displacements
\begin{equation}
  \label{eq:HWKernelDisp}
  \op{w} (x,  p) = \op{D} (x, p) \,\op{w} (0, 0) \,
  \op{D}^\dagger (x, p) \, ,
\end{equation}
where $\op{w}(0,0)=\int_{\mathbb{R}^{2}} \op{D}( x, p) \, dx dp = 2  \op{P}$,
and 
 \begin{equation}
  \label{eq:2}
  \op{P} = \int_{\mathbb{R}} | x \rangle \langle - x | \, dx = 
  \int_{\mathbb{R}} | p \rangle \langle - p | \, d p = 
 (-1 )^{\op{a}^{\dagger} \op{a}} 
\end{equation}
is the parity operator.

Let $\op{A}$ be an arbitrary operator acting on the Hilbert space of
the system.  Using the Stratonovich-Weyl quantizer we can associate to
$\op{A}$ a function $a(x, p)$ representing the action of the
corresponding dynamical variable in phase space. In fact, this is
known as the Wigner-Weyl map and is given by~\cite{Brif:1998qf}
\begin{equation}
  \label{eq:Asy}
  a(x, p)  =   \Tr [ \op{A} \,\op{w} (x, p) ] \, . 
\end{equation}
The function $a (x, p)$ is the symbol of the operator
$\op{A}$. Conversely, we can reconstruct the operator from its symbol
through
\begin{equation}
  \label{eq:WigWeyl}
  \op{A} = \frac{1}{(2\pi)^{2}} \int_{\mathbb{R}^{2}} a (x, p) \, 
  \op{w} (x, p ) \, dx dp \, .
\end{equation}

In this context, the Wigner function is nothing but the symbol of the
density matrix $\op{\varrho}$, and therefore
\begin{eqnarray}
  \label{eq:Wigcan}
  &  W_{\op{\varrho}}(x, p)  =   \Tr [ \op{\varrho} \,\op{w} (x, p) ]
  \, , &  \nonumber \\ 
  & &  \label{eq:HWWignerDef} \\
  & \op{\varrho}   =   \displaystyle 
  \frac{1}{(2\pi)^{2}} \int_{\mathbb{R}^{2}} \op{w}(x,p)
  W_{\op{\varrho}} (x, p) \, dx dp \, . &
  \nonumber
\end{eqnarray}
For a pure state $| \Psi \rangle$, it simplifies
\begin{equation}
  \label{eq:1}
  W_{|\Psi \rangle} (x, p) = \frac{1}{4\pi}  \int_{\mathbb{R}} 
  \Psi^{\ast} (x - x^{\prime} ) \, \Psi (x + x^{\prime} )  \, 
  \exp (i 2 p x^{\prime} ) \,  dx^{\prime} \, ,
\end{equation}
which is, perhaps, the most convenient form for actual calculations.

The Wigner function defined in (\ref{eq:HWWignerDef}) fulfills all the
basic properties required for any good probabilistic description.
First, due to the Hermiticity of $\op{w}(x, p)$, it is real for
Hermitian operators. Second, the probability distributions for the canonical
variables can be obtained as the marginals
\begin{equation}
  \label{eq:HWProps2}
  \int_\mathbb{R}  W_{\op{\varrho}} (x, p) \, dp =   
  \langle x | \op{\varrho} | x \rangle   \, ,
  \quad
  \int_\mathbb{R} W_{\op{\varrho}}(x, p) \, dx = 
  \langle p | \op{\varrho} | p \rangle  \, .
\end{equation}
Third, $W_{\op{\varrho}} (x, p)$ is translationally covariant, which
means that for the displaced state $\op{\varrho}^\prime =
\op{D}(x^{\prime}, p^{\prime})$ $\op{\varrho} \, \op{D}^\dagger
(x^{\prime}, p^{\prime})$, one has
\begin{equation}
  \label{eq:HWProps3}
  W_{\op{\varrho}^\prime} (x, p) = W_{\op{\varrho}} (x -x^{\prime}, p -p^{\prime}) \, ,
\end{equation}
so that it follows displacements rigidly without changing its form,
reflecting the fact that physics should not depend on a certain choice
of the origin. 

Finally, the overlap of two density operators is proportional to the
integral of the associated Wigner functions:
\begin{equation}
  \label{eq:HWProps4}
  \Tr ( \op{\varrho} \,\op{\varrho}^{\prime} ) \propto
  \int_{\mathbb{R}^2} W_{\op{\varrho}} (x, p)  W_{\op{\varrho}^{\prime}} (x, p)  \, dx dp \, .
\end{equation}
This property (known as traciality) offers practical advantages,
since it allows one to predict the statistics of any outcome, once the
Wigner function of the measured state is known.

The displacements constitute also a basic ingredient in the concept of
coherent states. If we choose a fixed normalized reference state 
$ | \Psi_{0}\rangle $, we have~\cite{Perelomov:1986kl}
\begin{equation}
  | x, p \rangle = \op{D} ( x, p) \, | \Psi_{0} \rangle \, ,  
  \label{eq:defCS}
\end{equation}
so they are parametrized by phase-space points. These states have a
number of remarkable properties inherited from those of $\op{D} (x,
p)$. In particular, $\op{D} (x, p)$ transforms any coherent state in
another coherent state:
\begin{equation}
  \op{D} ( x^{\prime},  p^{\prime} ) \, | x, p \rangle
  = \exp[i (x^{\prime} p - p^{\prime} x )/2] \,
  | x + x^{\prime}, p + p^{\prime} \rangle \, .
  \label{eq:comcoh}
\end{equation}
The standard choice for the fiducial vector $| \Psi_{0} \rangle$ is
the vacuum $|0 \rangle $;  which has quite a number of relevant
properties.

\section{Phase-space picture of a two-dimensional harmonic oscillator}
\label{Sec:2DHO}

Next, we analyze the superposition of two oscillators in orthogonal
directions, say $x$ and $y$, with momenta $\op{p}_{x}$ and
$\op{p}_{y}$, respectively. The corresponding complex amplitudes
$\op{a}_{x}$ and $\op{a}_{y}$ fulfill
$[\op{a}_{j},\op{a}_{k}^{\dagger}] =\delta_{jk} \op{\openone}$ ($j,
k\in \{x,y\}$).  Since these oscillators are kinematically independent
(i.e., they play the role of modes for the problem), the total system
is represented by the product of the corresponding kernels
\begin{equation}
  \op{w}(x, p_{x}; y, p_{y} ) = 
  \op{w} (x, p_{x} )\,\op{w}(y, p_{y} ) \, . 
  \label{eq:prodw1}
\end{equation}
The information is thus encoded in the four real  variables $(x, p_{x})$
and $(y, p_{y})$. The resulting Wigner function $W(x,p_{x};
y, p_{y})$ is informationally complete, but it is hard to grasp any
physical flavor from it. In particular, it cannot be plotted (which is
always a major advantage when depicting complex phenomena) and one
must content oneself with sections of $W(x, p_{x}; y, p_{y})$, which
illustrate only partial aspects~\cite{Singh:2007uq}. 

Because we are interested in elaborating on the behavior of OAM, which
mostly appears when cylindrical symmetry is present, we make the
change from Cartesian $(x, y)$ to polar $(r, \varphi)$ coordinates:
\begin{equation}
  r=\sqrt{x^{2}+y^{2}} \,, 
  \qquad
  \varphi=\arctan(y/x) \,.
\end{equation}
Simultaneously, we change from $(p_{x},p_{y})$ to
\begin{equation}
  p_{r}=\frac{1}{r}(xp_{x}+yp_{y}) \, , 
  \qquad 
  \ell=x p_{y} - y p_{x} \,,
\end{equation}
where $p_{r}$ is the radial momentum and $\ell$ is the OAM.  This
transition from Cartesian to polar coordinates is not smooth at the
origin and needs qualification because it takes from a contractible
space to one which is not contractible. This lies at the root of the
problems appearing when dealing with angle
variables~\cite{Carruthers:1968qq,Garrison:1970kx,Lerner:1970qd,
  Newton:1980rt,Leacock:1987fr,Ellinas:1991eu,Ma:1991la,Luis:1993mz}. In
quantum optics there are, however, a number of ways to bypass this
drawback~\cite{Susskind:1964wd,Paul:1974rw,Shapiro:1984ys,Barnett:1989th,
  Popov:1991gd,Noh:1992pb,Hradil:1992dk,Freyberger:1993bv}. In the
same vein, the radial momentum $p_{r}$ is singular at the origin,
which reflects a classical symptom of quantum
illness~\cite{Zhu:1993kb}, for such an operator is not selfadjoint
(nor has selfadjoint
extensions)~\cite{Liboff:1998kx,Galindo:1991fk,Paz:2001vn}. Precisely,
the use of Wigner-Weyl kernels alleviate these problems arising in a
direct quantization. However, we brush aside these mathematical
subtleties and move on to find a suitable solution for our problem.

Using the explicit form (\ref{eq:HWKernelDisp}) for each orthogonal
oscillator and after disentangling the exponentials, we can rewrite
(\ref{eq:prodw1}) in the equivalent way
\begin{eqnarray}
  \label{eq:KernelWithSines}
  \op{w} (r,p_{r};\varphi,\ell) & = & 4(-1)^{\op{N}} \nonumber \\
  & \times & \exp\left[-2\cos\varphi 
    (\alpha_{r}\hat{a}_{x}^{\dagger} -  \alpha_{r}^{*}\hat{a}_{x})\right]
  \nonumber \\ 
  & \times & \exp\left[2i\lambda_{r}\sin\varphi 
    (\hat{a}_{x}^{\dagger} +  \hat{a}_{x})\right ] 
  \nonumber \\
  & \times & \exp\left[-2\sin\varphi 
    ( \alpha_{r}\hat{a}_{y}^{\dagger} - \alpha_{r}^{*}\hat{a}_{y})\right]
  \nonumber \\ 
  & \times & \exp\left[-2i\lambda_{r}\cos\varphi 
    (\hat{a}_{y}^{\dagger} +  \hat{a}_{y})\right] \,,
\end{eqnarray}
where
\begin{equation}
\label{N}
\op{N}=\op{a}_{x}^{\dagger}\op{a}_{x}+\op{a}_{y}^{\dagger}\op{a}_{y}
\end{equation}
is the total number of excitations and we have denoted
$\alpha_{r} =(r+ip_{r})/\sqrt{2}$ and $\lambda_{r} = \ell / (\sqrt{2}r)$. 

The structure of this kernel suggests the use of the rotated operators
\begin{equation}
  \label{eq:rotated}
  \op{a}_{+}=\frac{1}{\sqrt{2}}(\op{a}_{x}-i\op{a}_{y})\, , \qquad
  \op{a}_{-}= \frac{1}{\sqrt{2}}(\op{a}_{x}+i\op{a}_{y})\, , 
\end{equation}
in terms of which the OAM operator reads as
\begin{equation}
  \label{N}
  \op{L}=\op{a}_{+}^{\dagger}\op{a}_{+}
  -\op{a}_{-}^{\dagger}\op{a}_{-} \,  .
\end{equation}
In this way, we can interpret $\ell$ as the difference of quanta with
opposite chirality. Note that the form of $\op{N}$ and $\op{L}$
suggests that the boson operators $\op{a}_{x}$ and $\op{a}_{y}$
furnish a Jordan-Schwinger representation for the problem at hand
[much in the same way as the original oscillator construction for
SU(2)], which can be justified on very general
grounds~\cite{Chaturvedi:2006kl}. On the other hand, such a
representation should not come as a surprise, for it is well known
that any three-dimensional Lie algebra (as the one we are dealing with
here) can be realized in terms of creation and annihilation operators
of two orthogonal oscillators~\cite{Gracia-Bondia:2002tg}.

By noticing that $e^{-i\varphi\op{L}}\,\op{a}_{\pm}\, e^{i\varphi\op{L}}=
\op{a}_{\pm}e^{\pm i\varphi}$, we can recast the Wigner kernel
\eqref{eq:KernelWithSines}  as  the  displaced version 
\begin{equation}
\hat{w} (r, p_{r};\varphi, \ell ) =  
 e^{-i\varphi\hat{L}} \,   \hat{w}(r,p_{r};\ell) \, e^{i\varphi\hat{L}}\, , 
\label{eq:6}
\end{equation}
 with 
\begin{eqnarray}
\op{w} (r, p_{r},\ell) &  = &
4(-1)^{\op{N}} \nonumber \\
& \times & \exp[2i\lambda_{r} (\op{p}_{+}-\op{p}_{-})]  \nonumber \\  
 & \times & \exp[-\sqrt{2} i  p_{r} (\op{x}_{+}+\op{x}_{-})] \nonumber \\
& \times & \exp[\sqrt{2}i r(\op{p}_{+}+\op{p}_{-})]e^{-2ip_{r}r} \, ,
\end{eqnarray}
and  we have introduced the corresponding quadratures for the rotated
amplitudes
\begin{equation}
  \op{x}_{\pm}=\frac{1}{\sqrt{2}}(\op{a}_{\pm}+\op{a}_{\pm}^{\dagger})\,,
  \qquad
  \op{p}_{\pm}=\frac{1}{\sqrt{2} i}(\op{a}_{\pm}-\op{a}_{\pm}^{\dagger})\,.       
\end{equation} 

The radial momentum $p_{r}$ plays no relevant role in the dynamics, so
it seems entirely reasonable to integrate over this variable. To
evaluate the resulting kernel we use an entangled state basis
$|\xi\rangle$ (the properties of these states are briefly reviewed in
Appendix~\ref{EPR}), such that
\begin{equation}
  (\op{a}_{+}+\op{a}_{-}^{\dagger})|\xi\rangle=\xi|\xi\rangle\,,
  \qquad
  (\op{a}_{+}^{\dagger}+\op{a}_{-})|\xi\rangle=\xi^{\ast}|\xi\rangle\,.   
\end{equation}
The calculations are lengthy and the details are sketched in
Appendix~\ref{Wigker}.  The final result for the Wigner
function for a pure state $| \Psi \rangle$ turns out to be remarkably
simple:
\begin{equation}
  \begin{aligned}
    \label{eq:cen}
    W_{|\Psi \rangle} (r, \varphi, \ell) = 4\int_{\mathbb{R}}
    \Psi^{\ast} (r-i r^{\prime}, \varphi) \, \Psi (r +
    ir^{\prime},\varphi) \exp( i 2 \ell r^{\prime} /r ) \,
    dr^{\prime}\,,
  \end{aligned}
\end{equation}
 which is the central result of this work. Here $\Psi(r,\phi)$ denotes
the wave function of $|\Psi \rangle$ in the entangled representation; i.e.,
\begin{equation}
\Psi ( r, \varphi)  =\langle \xi|e^{i\varphi\op{L}} | \Psi\rangle = 
 \langle\xi e^{-i\varphi} |\Psi\rangle \, . 
 \label{eq:psi} 
\end{equation}
Notice that the similarity with the single-mode Wigner function
(\ref{eq:1}) is manifest. Obviously, the marginal over the radial
variable
\begin{equation}
  \label{eq:Wfil}
  W (\varphi, \ell) = \int W (r, \varphi, \ell ) \, dr \, , 
\end{equation}
contains complete information about the pair angle-OAM and can be
constructed from first principles~\cite{Rigas:2010kx}. 

\section{Examples}
\label{Sec:ex}

\begin{figure}[b]
 \includegraphics[width=0.50\columnwidth]{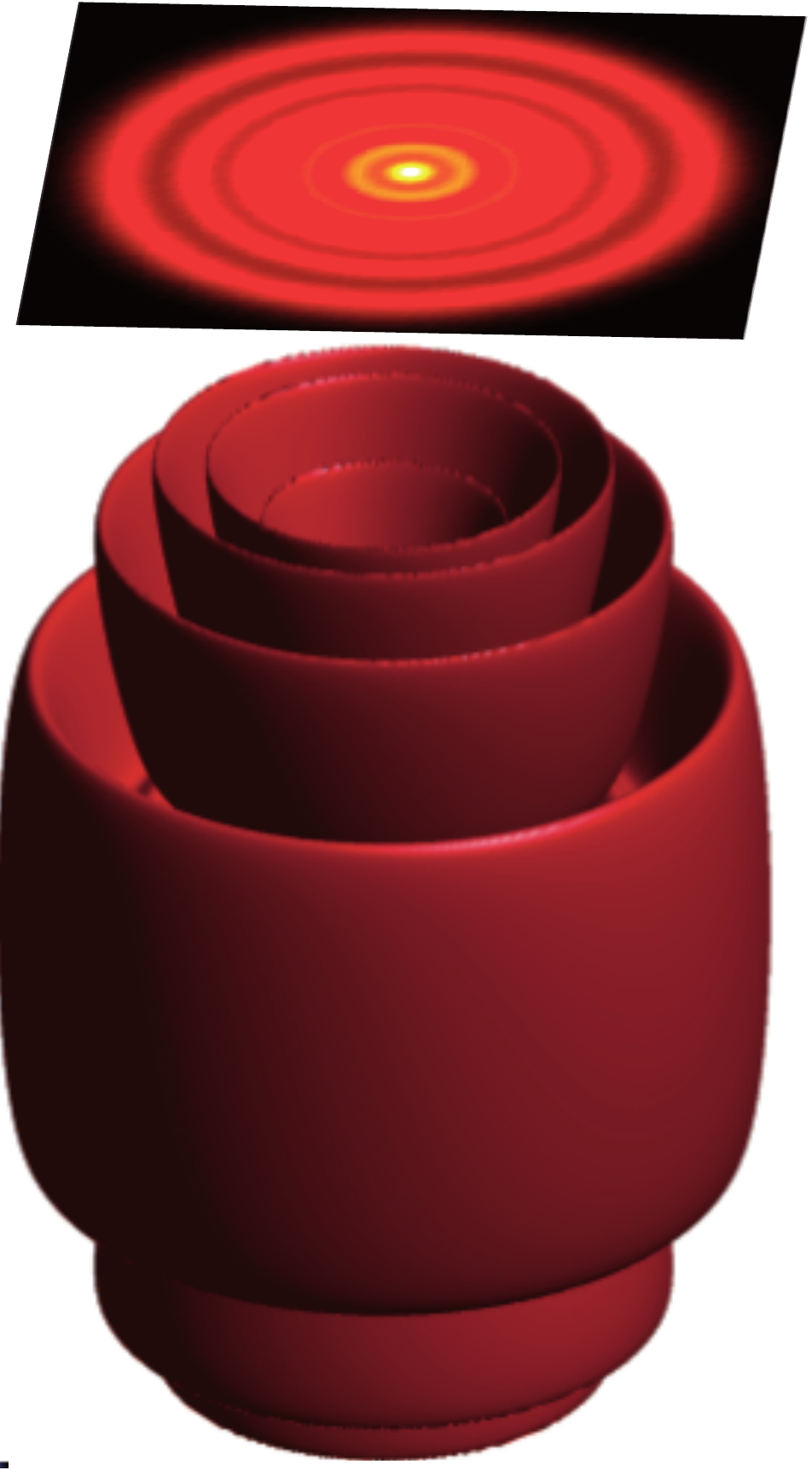}
\label{Fig:OAM0}
  \caption{Isocontour surface of the level $1/e$ from the maximum of
    the Wigner function  $W (r, \varphi, \ell)$ for an eigenstate of the
    OAM $ | \ell_{0} \rangle$. At the top, we show a density plot of a
    section of that surface by the plane $\ell = 0.$}
  \end{figure}

To gain further insight into this formalism, we  work out
(\ref{eq:cen}) for several states of interest. First, we look for the
case of a simultaneous eigenstate of both the total number of
particles and the orbital angular momenta $ | N, \ell_{0} \rangle $, viz
\begin{equation}
  \label{eq:2}
  \op{N} |N,\ell_{0} \rangle = N |N,\ell_{0} \rangle \, ,   
\qquad
\op{L} |N,\ell_{0}\rangle=\ell_{0} |N,\ell_{0}\rangle \, .
\end{equation}
Using the entangled representation, it is easy to check
that  
\begin{eqnarray}
\Psi_{|N,\ell_{0} \rangle} ( r, \varphi) & = & \frac{e^{-|\xi|^2/2}}
 {\sqrt{\left ( \frac{N -|\ell_{0}|}{2} \right ) ! 
\left  ( \frac{N +|\ell_{0}|}{2} \right ) !}}  \nonumber \\
& \times & 
H_{\frac{N - |\ell_{0}|}{2},\frac{N + |\ell_{0}|}{2}}(\xi e^{-i\varphi}, \xi^{\ast} e^{i\varphi}) \, , 
\end{eqnarray}
where $H_{m,n} (\lambda,\lambda^{\ast})$ stands for the two-variable
Hermite polynomial. In terms of the generalized Laguerre polynomials
$L_{p}^{\ell} (x)$, this reduces to 
\begin{equation}
  \label{eq:7}
  \Psi_{|N,\ell_{0} \rangle} (r, \varphi) = C_{N,\ell_{0}}
  e^{-\frac{1}{2} r^2}  r^{|\ell_{0}|}  
  L_{\frac{N -|\ell_{0}|}{2}}^{|\ell_{0}|} (r^2)  e^{-i\ell_{0} \varphi} \, ,
\end{equation}
where $C_{N,\ell_{0}}$ is a normalization constant. This wave function is
very reminiscent of the standard Laguerre-Gauss modes employed in
classical optics. The associated Wigner function is
\begin{eqnarray}
W_{|N,\ell_{0} \rangle}  (r, \varphi, \ell ) & = & 
4 | C_{N,\ell_{0}}|^{2} \int_{\mathbb{R}} 
 (r + i r^{\prime})^{2 |\ell_{0} |}  [ L^{\ell_{0}}_{\frac{N - |\ell_{0}|}{2}}
 (r^2+r'^2 ) ]^{2} \nonumber \\
& \times & \exp[ - (r^2+r^{\prime 2} + 2 i \ell r^{\prime}/r )]   \, dr^{\prime} \, .
\end{eqnarray}
This integral can be computed in a closed way, although the expression
is involved enough to be of practical use. If we sum over $N$, we
get the state
\begin{equation}
  \label{eq:9}
  | \ell_{0} \rangle = \sum_{N} \frac{1}{\sqrt{N+1}} 
 |N, \ell_{0}  \rangle \, . 
\end{equation}
In Fig.~1 we have plotted an isocontour surface corresponding to $W_{|
  \ell_{0} \rangle} (r, \varphi, \ell ) = $ constant, for $\ell_{0} =
0$. We clearly appreciate quite a rich radial structure. At the top of
the surface, we also include a density plot of a section by the plane
$\ell = 0$, displaying the characteristic rings of the Laguerre
modes.  We recall that the standard Wigner function for the pair
angle-OAM simplifies in this case to
\begin{equation}
  \label{eq:ExampleOAMstaleEll}
  W_{| \ell_0 \rangle} (\ell,\phi) = \frac{1}{2\pi} 
  \delta_{\ell, \ell_0} \, , 
\end{equation}
which is flat in $\phi$ and the integral over the whole phase space
gives the unity, reflecting the normalization of $|\ell_0\rangle$.  We
can recognize the amount of information lost in this approach when
compared with $W( r, \varphi, \ell )$.  A similar procedure can
be used for the case of the eigenstates of the angle $| \varphi_{0}
\rangle$.

\begin{figure}
  \includegraphics[width=0.90\columnwidth]{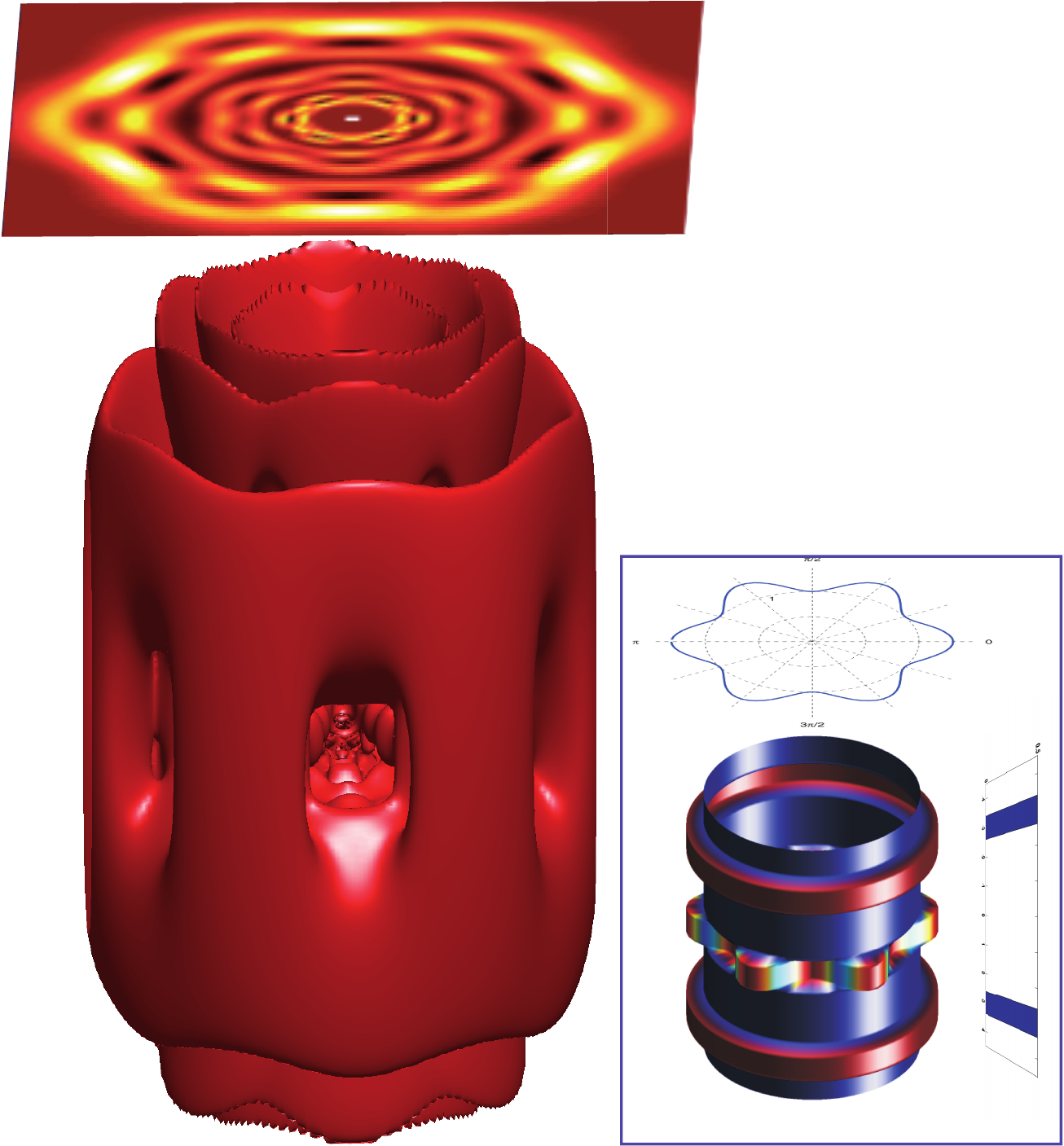}
  \label{Fig:tornillo}
  \caption{Isocontour surface of the level $1/e$ from the
      maximum of the Wigner function $W (r, \varphi, \ell)$ for the
      superposition state in Eq.~(\ref{eq:SuperDef}), with $\ell_2 =
      -3$ and $\ell_1 = 3$. At the top, we show a density plot of a
      section of that surface by the plane $\ell = 3$. In the inset
      we show the standard Wigner function $W(\varphi , \ell)$ for
      this state, as well as the associated marginals.}
  \end{figure}

  As our second example, we address the superposition
  \begin{equation}
    \label{eq:SuperDef}
    | \Psi \rangle = \frac{1}{\sqrt{2}} 
    (|\ell_{1} \rangle   + e^{i\phi_0} | \ell_{2} \rangle )
  \end{equation}
  of two angular-momentum eigenstates with a relative phase $e^{i
    \phi_0}$. The resulting features are nicely illustrated in
  Fig.~2. The state $|\Psi \rangle $ is plotted for $\ell_{2} = -3$
  and $\ell_{1} = 3$.  Changing the relative phase $\phi_0$ results in
  a global rotation of the cylinder. Again a rich radial structure can
  be appreciated. The ``holes'' in the isosurface correspond to points
  for which the Wigner function takes on negative
  values~\cite{Rigas:2010rt}, as can be appreciated in the inset,
  where we draft the corresponding Wigner function $W (\varphi, \ell)$
  for this state.

\section{Concluding remarks}
\label{Sec:Con}

In summary, we have shown how to extend in a consistent way all
the techniques developed for a continuous-variable phase space to the case
of angle and angular momentum, including significant information about
the radial variable. While we have not left aside the
mathematical details, our main emphasis has been on presenting a
simple and useful toolkit that any practitioner in the field should
master. In our view, far from being an academic curiosity, the ideas
expressed here have a wide range of potential applications in
numerous hot topics in which OAM plays a key role. 

\begin{acknowledgments}
  The inspiring ideas in this paper originated after many discussions
  with Prof. W. Schleich. Over the years, they have been further
  developed and completed with questions, suggestions, criticism, and
  advice from many colleagues. Particular thanks for help in various
  ways go to A. G. Barriuso, B.-G. Englert, J. C. Gallego, H. de Guise, and H. Kastrup.

  The work was supported by the EU FP7 (Grant Q-ESSENCE), the Spanish
  DGI (Grant FIS2011-26786), the Mexican CONACyT (Grant 106525), the
  Czech Ministry of Education (Project MSM6198959213), and the Czech
  Ministry of Industry and Trade (Project FR-TI1/364).
\end{acknowledgments}

\appendix

\section{Entangled-state representation}

\label{EPR}

For the two modes $\pm$ defined in Eq.~(\ref{eq:rotated}), the Fock space is spanned
by the vectors 
\begin{equation}
|n_{+},n_{+}\rangle=\frac{(\op{a}_{+}^{\dagger})^{n_{+}}
  (\op{a}_{-}^{\dagger})^{n_{-}}}{\sqrt{{n_{+}}!{n_{-}}!}}|0,0\rangle\,,
\end{equation}
 where $|0,0\rangle$ is the two-mode vacum. Then one can immediately
check that the vectors~\cite{Fan:2002kx,Fan:2010ys}
\begin{equation}
\begin{aligned}|\xi\rangle & =\exp\left[-\frac{1}{2}|\xi|^{2}+
    \xi\op{a}_{+}^{\dagger}+\xi^{\ast}\op{a}_{-}^{\dagger}-
    \op{a}_{+}^{\dagger} \op{a}_{-}^{\dagger}\right]|0,0\rangle\,,\\
|\eta\rangle & =\exp\left[-\frac{1}{2}|\eta|^{2}+\eta
  \op{a}_{+}^{\dagger}-\eta^{\ast}\op{a}_{-}^{\dagger}+
  \op{a}_{+}^{\dagger}\op{a}_{-}^{\dagger} \right]|0,0\rangle\,,
\end{aligned}
\end{equation}
 are indeed eigenstates of the following operators 
\begin{equation}
\begin{aligned}
\op{x}_{+}-\op{x}_{-}|\eta\rangle &
=\sqrt{2} \, {\rm     Re}(\eta)|\eta\rangle\,, \quad
\op{p}_{+}+\op{p}_{-}
  |\eta\rangle=  \sqrt{2}\,{\rm Im}(\eta)|\eta\rangle\,,\\ 
\\
 \op{x}_{+}+\op{x}_{-}|\xi\rangle & 
=\sqrt{2} \, {\rm Re}(\xi)|
\xi\rangle\,,\quad
\op{p}_{+}-\op{p}_{-} |\xi\rangle=  
\sqrt{2} \,{\rm
  Im}(\xi)|\xi\rangle\,, 
\end{aligned}
\end{equation}
 where $\hat{x}_{\pm}$ and $\hat{p}_{\pm}$ are the quadrature operators
associated to the modes $\pm$. This shows that these states are the
continuous-variable version of the original Einstein-Podolsky-Rosen
states~\cite{Einstein:1935zr}.

Using the technique of integration within an ordered product of
operators~\cite{Fan:1987ve},  we can prove the orthogonal property and
completeness relation  
\begin{equation}
\langle\eta^{\prime}|\eta\rangle=\pi\delta^{(2)}(\eta-\eta^{\prime})\,,\qquad
\frac{1}{\pi}\int d^{2}\eta\,|\eta\rangle\langle\eta|=\op{\openone}\,,
\end{equation}
 and an analogous one for $\xi$. In fact, one can also check that
\begin{equation}
\langle\xi|\eta\rangle=\frac{1}{2}\exp[(\xi\eta^{\ast}-\xi^{\ast}\eta)/2]\,.
\end{equation}
 We observe also that if we use the Shapiro-Wagner angle operator~\cite{Shapiro:1984ys}
\begin{equation}
\op{E}  = \sqrt{\frac{\op{a}_{+}+\op{a}_{-}^{\dagger}}
 {\op{a}_{+}^{\dagger}+   \op{a}_{-}}}   \,,
\end{equation}
then 
\begin{equation}
\op{E}| \xi\rangle=
\sqrt{\frac{\op{a}_{+}+\op{a}_{-}^{\dagger}}{\op{a}_{+}^{\dagger}+ 
    \op{a}_{-}}}|\xi\rangle=\sqrt{\frac{\xi}{\xi^{*}}}|\xi\rangle=e^{i\phi}|
\xi\rangle\,,
\end{equation}
 so these states have a well-defined angle.

If we recall that the two-variable Hermite polynomials, 
defined as~\cite{Dodonov:1994qf}
\begin{equation}
H_{m,n}(\lambda,\lambda^{\ast}) =
\sum_{\ell=0}^{\min(m,n)}\frac{m!n!}{\ell!(m-\ell)!(n-\ell)!} 
 (-1)^{\ell}\lambda^{m-\ell}{\lambda^{\ast}}^{n-\ell}\,,
 \label{eq:2varHer}
\end{equation}
 have the generating function 
\begin{equation}
\sum_{m,n}^{\infty}\frac{t^{m}{t^{\prime}}^{n}}{m!n!}H_{m,n}(\lambda,\lambda^{*})=
\exp(-tt^{\prime}+t\lambda+t^{\prime}\lambda^{\ast})\,,
\end{equation}
 by simple inspection we note that 
\begin{equation}
\label{eq:VVD}
\begin{aligned}
|\eta\rangle & = \exp (- |\eta|^{2}/2)
\sum_{n_{+},n_{-}}\frac{(-1)^{n_{-}}}
{\sqrt{n_{+}!n_{-}!}}
H_{n_{+},n_{-}}(\eta,\eta^{*})|n_{+},n_{-}\rangle \,,\\ 
|\xi\rangle & = \exp (- |\xi|^{2}/2)  \sum_{n_{+},n_{-}}  
\frac{1}{\sqrt{n_{+}!n_{-}!}}
H_{n_{+},n_{-}}(\xi,\xi^{\ast})|n_{+},n_{-}\rangle\,, 
\end{aligned}
\end{equation}
 which constitute a compact expression of these entangled vectors
in the Fock basis.

\section{Evaluating the Wigner-Weyl kernel}

\label{Wigker}

Our task here is to evaluate the kernel (\ref{eq:6}) and then
integrate over the variable $p_{r}$. Using the properties of the
entangled states in the previous Appendix, we can write
\begin{equation}
\begin{aligned}
\op{w}(r,\ell)= & \int dp_{r}\ \op{w}(r,p_{r},\ell)\\ 
= & \frac{1}{\pi}(-1)^{\op{N}}\int  d^{2}\eta d^{2}\xi\,
|\eta\rangle\langle\xi|\exp[(\xi\eta^{\ast}-\xi^{\ast}\eta)/2]\\ 
\times & \exp[\sqrt{2}\lambda_r(\xi-\xi^{\ast})] \exp[ r
(\eta-\eta^{\ast})]  \delta\left[r-\mathrm{Re}(\xi) \right]\,.
\end{aligned}
\end{equation}
To simplify as much as possible what follows, we assume pure
states, for which 
\begin{equation}
W(r,\varphi,\ell)=\langle\Psi|\op{w}(r,\varphi,\ell)|\Psi\rangle=
\langle\Psi|e^{-i\varphi\op{L}}\op{w}(r,\ell)e^{i\varphi\op{L}}|\Psi\rangle\,. 
\end{equation}
 This is in fact a marginal of the Wigner function of
the problem. Next, we choose to expand
$e^{i\varphi\op{L}}|\Psi\rangle$ 
in the $|\xi\rangle$ basis. Taking into account the properties of
these states, we have 
\begin{equation}
\Psi(\xi,\varphi)=\langle\xi|e^{i\varphi\op{L}}| \Psi\rangle=\langle\xi
e^{-i\varphi}|\Psi\rangle\,. 
\end{equation}
 Therefore, we get 
\begin{equation}
\begin{aligned}W(r,\varphi,\ell) & =\frac{1}{\pi^{2 }}\int
  d^{2}\xi^{\prime} 
d^{2}\eta d^{2}\xi\,
  \langle\xi^{\prime}|(-1)^{\op{N}}|\eta\rangle \,
\delta\left[r-\mathrm{Re}(\xi)\right] \\  
 & \times\Psi^{\ast}(\xi^{\prime},\varphi)\Psi(\xi,\varphi)
 \exp[(\xi\eta^{\ast}-\xi^{\ast}\eta)/2]\\ 
 & \times\exp[\sqrt{2} \lambda_r (\xi-\xi^{\ast})]\exp[
 r(\eta-\eta^{\ast})]  \, .
\end{aligned}
\end{equation}

If we use the decomposition of these entangled states in terms of
double-variable Hermite polynomials in Eq.~(\ref{eq:VVD}), and we
recall that $H_{m,n}(\xi,\xi^{\ast})=H_{n,m}^{\ast}(\xi,\xi^{\ast})$,
then it is easy to check that $\langle\xi^{\prime}|(-1)^{\op{N}}|\eta\rangle=
\langle\eta|\xi^{\prime}\rangle$.  Consequently, we have
\begin{eqnarray} \nonumber 
& \displaystyle
\int d^{2}\eta\langle\eta|\xi^{\prime}\rangle\,
\exp[(\xi\eta^{\ast}-\xi^{\ast}\eta)/2+ r(\eta-\eta^{\ast})] & \\ 
& \displaystyle
=
4\pi^{2}\delta^{(2)}\left(\xi+\xi^{\prime} -
  2r\right)\,. 
& 
\end{eqnarray}
 Finally, if we perform the integral over $\xi^{\prime}$ using
this result we get 
\begin{equation}
\begin{aligned}W(r,\varphi,\ell) & =4\int d^{2}\xi\
  \Psi^{\ast}(2r-\xi,\varphi)\Psi(\xi,\varphi)\\ 
 & \times\exp[\sqrt{2}\lambda_r (\xi-\xi^{\ast})]\,
 \delta\left[ r-\mathrm{Re} (\xi) \right]\,. 
\end{aligned}
\end{equation}
By separating the differential $d^{2}\xi$ in real and imaginary
parts, after integrating over the real part $\mathrm{Re} (\xi)$ we get
the result~\eqref{eq:cen}.


%

\end{document}